\documentclass[twocolumn,aps,prc,superscriptaddress,showpacs,floatfix]{revtex4}

\usepackage{graphicx}

\begin{document}
\title{Heavy Quark Production from Jet Conversions in a Quark-Gluon Plasma}

\author{W. Liu}
\affiliation{Cyclotron Institute and Physics Department, Texas A$\&$M
University, College Station, Texas 77843-3366}
\author{R. J Fries}
\affiliation{Cyclotron Institute and Physics Department, Texas A$\&$M
University, College Station, Texas 77843-3366}
\affiliation{RIKEN/BNL Research Center, Brookhaven National Laboratory,
Upton, NY 11973}

\begin{abstract}
Recently, it has been demonstrated that the chemical composition of
jets in heavy ion collisions is significantly altered compared
to jets in the vacuum. This signal can be used to probe
the medium formed in nuclear collisions. In this study we investigate the
possibility that fast light quarks and gluons can convert to heavy quarks when
passing through a quark gluon plasma. We study the rate of
light to heavy jet conversions in a consistent Fokker-Planck framework and
investigate their impact on the production of high-$p_T$ charm and bottom
quarks at RHIC and LHC.
\end{abstract}

\pacs{12.38.Mh;24.85.+p;25.75.-q}

\maketitle

\section{Introduction}

In heavy ion collisions, energetic jets are expected to lose a significant
amount of energy when passing through the surrounding quark gluon plasma (QGP)
\cite{Baier:1996kr,wang, gyulassy, wiedemann}. This jet quenching
phenomenon leads to a large suppression of the yield of high-$p_T$ hadrons
in Au+Au collisions compared to the yield expected from a superposition of
$p+p$ collisions \cite{adcox,adler1} at the same center-of-mass energy.
Induced gluon radiation was expected to be the largest contributor to the
momentum degradation of the leading jet parton. However, experimental results
on the quenching of heavy flavors \cite{adare} has led to new efforts
to revisit elastic energy loss \cite{mustafa, wicks, djordjevic1}
and non-perturbative mechanisms \cite{hees}. Jet quenching measurements
deliver important information about the medium in terms of the average
squared momentum transfer to the jet per unit path length,
$\hat q = \mu^2/\lambda$.
Besides quenching as a manifestation of energy loss of the leading
high-$p_T$ parton, the chemical composition of a jet can also change
through flavor changes of the leading parton. This is an unavoidable
consequence of collisions of the high-$p_T$ parton with thermal partons.
The probability of conversion is related to the path length through the
medium and the collisional conversion widths. Thus by tracking the flavor
of jets (defined as the flavor of the leading parton) throughout their
interaction with the medium, transport properties of the medium which are
independent of and complementary to $\hat q$ can be obtained. Eventually,
the flavor changes will be reflected by the hadron composition of the jet.

Flavor changing processes were first studied in the context of light
quarks and gluons \cite{weiliu1}, and parton to photon conversions
\cite{Fries:2002kt}. More recently, we unified these two processes
and discussed new observables which could be measured at the Relativistic
Heavy Ion Collider (RHIC) or the Large Hadron Collider (LHC) \cite{weiliu2}.
In Ref.\ \cite{weiliu1} it was found that conversions of light quarks to
gluons could help solve the puzzle of very similar nuclear modification
factors $R_{AA}$ for pions and protons observed by the STAR
experiment \cite{adams}. These seemed to be incompatible with a relative
energy loss of 9/4 for gluons and light quarks. However, as we discussed
in more detail in \cite{weiliu2}, the concept of a fixed flavor for
the leading jet parton is ill-defined in a medium. Conversions to real
or virtual photons are just other examples of what can happen to a
fast quark or gluon traversing a quark gluon plasma. Although much rarer
than conversions to strongly interacting partons, it has been shown that
these photons can make a significant contribution to the direct photon
or dilepton spectrum at intermediate $p_T$.

In \cite{weiliu2} it was shown that strange hadrons might be the ideal probe
for flavor conversions at RHIC energies. This is due to the strong gradient
between the very small initial strange quark content of jets at RHIC,
and the almost full chemical equilibration of strangeness in the plasma.
Jets coupling to the medium will be driven toward chemical equilibrium and
thus kaons and $\Lambda$-hyperons have to be enhanced relative to pions and
protons at large $p_T$ when compared to naive expectations from $p+p$
collisions. Measurements of this enhancement could shed light on the mean
free path of fast gluons and light quarks in the medium. At LHC, the value
of strangeness as a high-$p_T$ probe is diminished through the large
initial strangeness content of jets at larger energies.
When translating leading particle flavor into hadrons it has to be kept
in mind that the hadron chemistry can also be altered through the increased
multiplicity in a jet cone interacting with the medium as studied
in \cite{Sapeta:2007ad}.

Two important questions emerge from the discussion of high-$p_T$ strangeness
enhancement at RHIC. First, can heavy quarks, in particular charm, play
a similar role as a probe for the average mean free path at LHC? And
secondly, is the suppression of charm found at RHIC compatible with
the mechanism proposed here which boosts the yields of rare particle
species at intermediate and high $p_T$?
In this Report, we answer these questions by studying the production of
high-$p_T$ heavy quarks from jet conversions in heavy ion collisions.
We compute the contributions from leading order (LO) processes in
perturbation theory multiplied by a $K$-factor. The heavy quarks can
be produced from the annihilation process $g+g\to Q+\bar Q$ that converts
a gluon jet and a thermal gluon to a pair of heavy quarks, and also through
the Compton processes $g(q)+Q\to Q+g(q)$ by transferring the momentum of the
incoming jet to that of a slow moving heavy quark in the medium. We follow the
method introduced in Ref.\ \cite{weiliu2} to study the propagation of
the high $p_T$ heavy quark in the expanding fireball within the framework
of a Fokker-Planck equation.

This work is organized as follows. In the next section we discuss the
transport coefficients of heavy quarks and the collisional widths for
the conversion of light partons to heavy quarks. We then proceed to
present numerical results for charm and bottom quarks at RHIC and LHC in
Sec.\ref{propagate} using a consistent model based on rate equations for
conversions and a Fokker-Planck equation for energy loss. Finally, a
summary is presented in Sec.\ \ref{summary}.

\section{Heavy Quark Scattering in a QGP}\label{trans_coeff}

First we discuss the propagation of heavy quarks in a partonic medium. It
can be described in the framework of a Fokker-Planck equation
\cite{benjamin,hees}
\begin{eqnarray}\label{fokkerplanck}
\frac{\partial f({\bf p},t)}{\partial t}=\frac{\partial}{\partial {\rm p_ i}}
\left[A_{\rm i}({\bf p})+\frac{\partial}{\partial {\rm p_ j}}B_{\rm ij}({\bf p})\right]f({\bf p},t)
\end{eqnarray}
where $f({\bf p},t)$ is the distribution function of heavy quarks.
The drag and diffusion coefficients are
\begin{eqnarray}\label{coefficient}
A_{\rm i}({\bf p})&=&{\rm p_i} \gamma({|\bf p|}),\nonumber\\
B_{\rm ij}({\bf p})&=&\left(\delta_{\rm ij}-\frac{{\rm p_ i} {\rm p_ j}}{|\bf p|^2}\right)B_0({|\bf p|})
  +\frac{\rm p_ip_j}{|{\bf p}|^2}B_1({|\bf p|}),
\end{eqnarray}
with
\begin{eqnarray}
\gamma({|\bf p|})&=&\langle 1\rangle -\frac{\langle{\bf p}\cdot{\bf p'}\rangle}{|\bf p|^2},\nonumber\\
B_0({|\bf p|})&=&\frac{1}{4}\left[\langle |{\bf p}|^2\rangle
   -\frac{\langle{(\bf p}\cdot{\bf p'})^2\rangle}{|{\bf p}|^2}\right],\nonumber\\
B_1({|\bf p|})&=&\frac{1}{2}\left[\frac{\langle{(\bf p}\cdot{\bf p'})^2\rangle}{|{\bf p}|^2}
  -2\langle{\bf p}\cdot{\bf p'}\rangle +{\bf p}^2\langle 1\rangle\right] \, .
\end{eqnarray}
Here ${\bf p}$ and ${\bf p'}$ label the momenta of the heavy quark before
and after each collision, respectively. We define the thermal averaging
$\langle Y(\bf p')\rangle$ in the elastic process $Q+g(q)\to Q+g(q)$ as
\begin{eqnarray}\label{average1}
&&\langle Y(\bf p')\rangle \nonumber\\
&=&\frac{1}{2E_p}\int\prod_{Q=p',q,q'}
\frac{d^3Q}{(2\pi)^32E_Q}\delta^{(4)}(p+q -p'-q')\nonumber\\
&&\times \frac{1}{\gamma_Q}(2\pi)^4|{\cal M}_{p+q\to p'+q'}|^2f(q)[1\pm f(q')]Y(\bf p').
\end{eqnarray}
where $\mathcal{M}$ is the scattering amplitude and $\gamma_Q=6$ is the spin and color degeneracy factor for heavy quark.

\begin{figure}[ht]
\includegraphics[width=2.0in,height=1.7in,angle=-90]{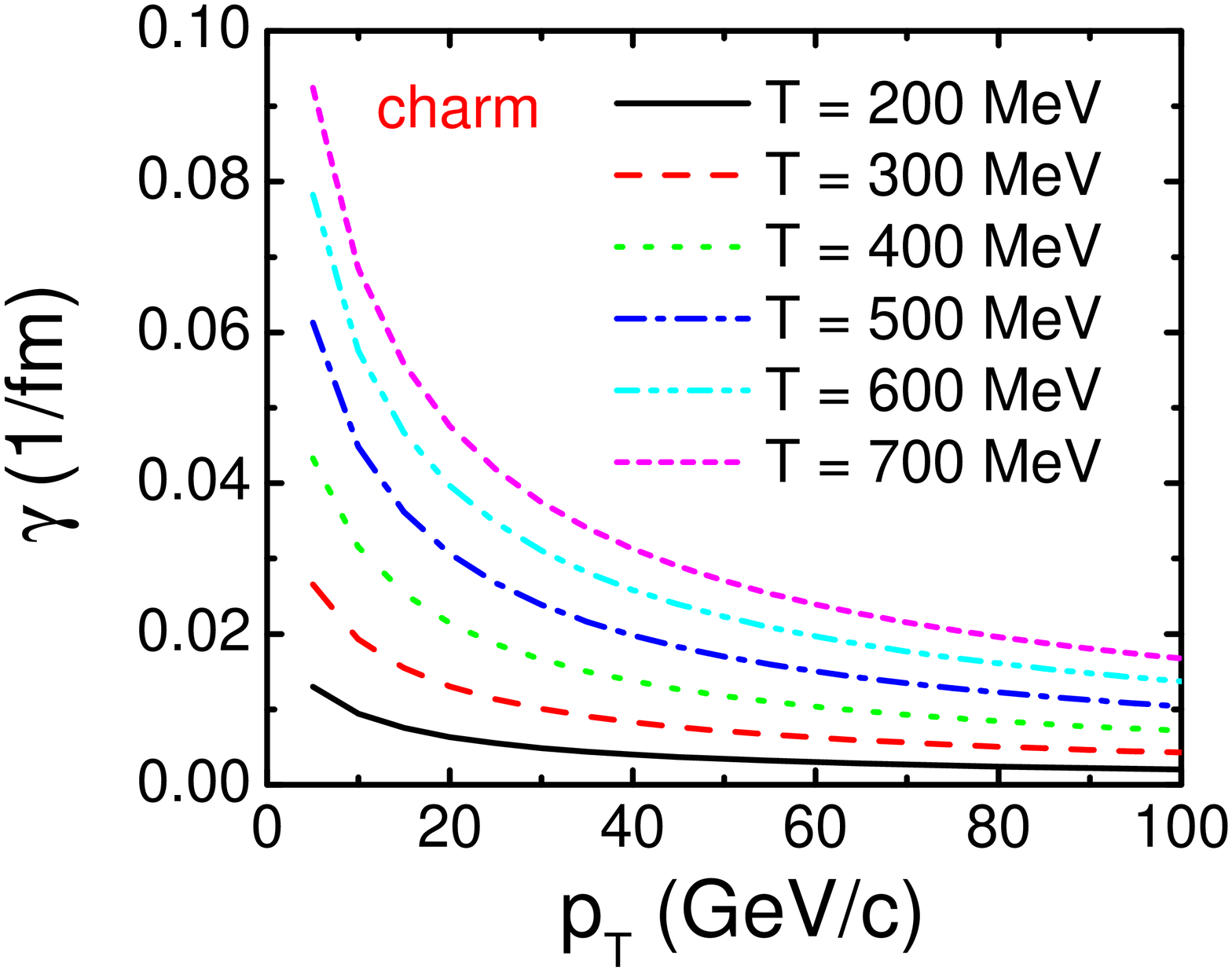}
\hspace{-0.5cm}
\includegraphics[width=2.0in,height=1.7in,angle=-90]{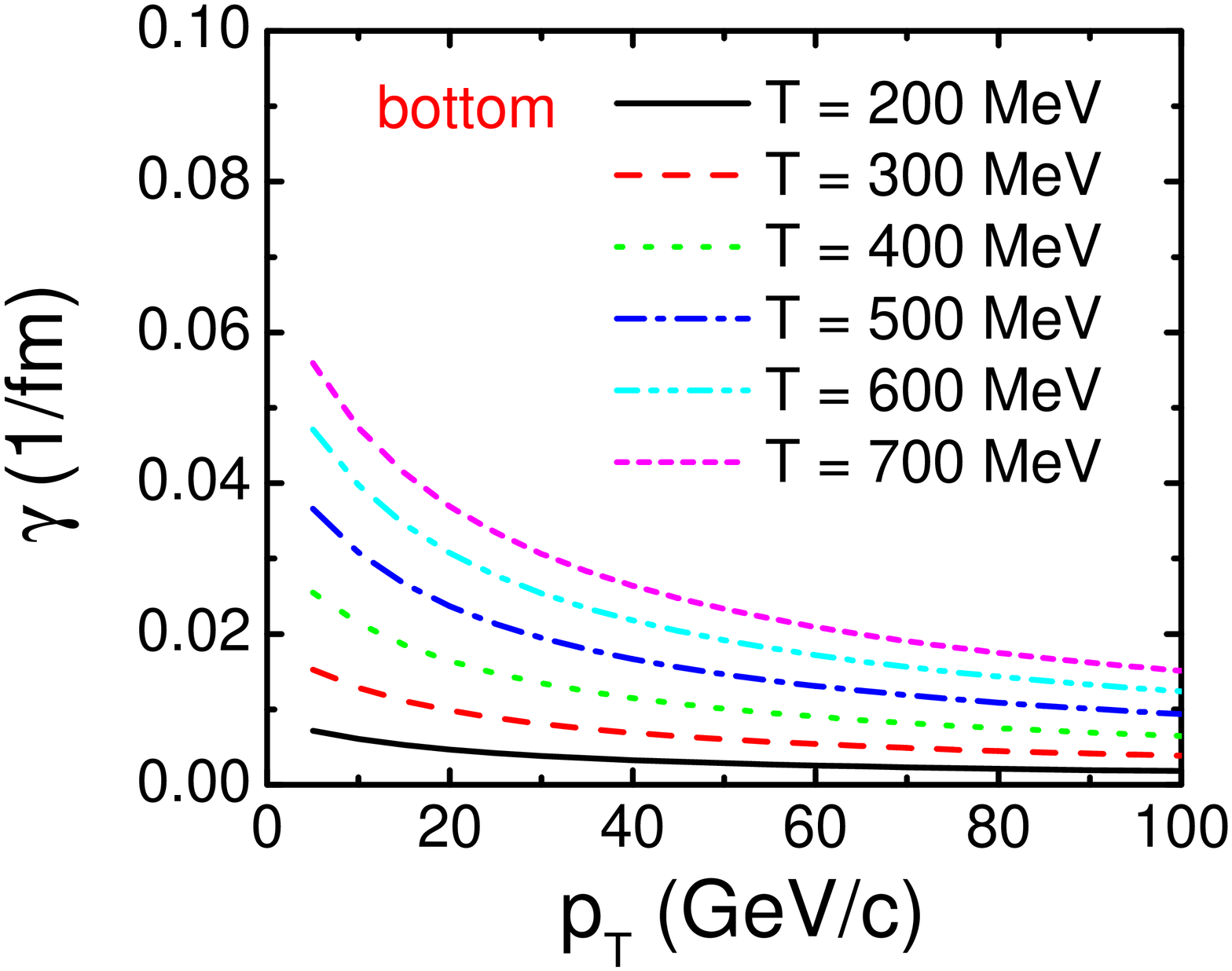}
\caption{(Color online) Transport coefficients $\gamma$ of charm (left) and bottom quarks (right) in a QGP as functions of transverse momentum $p_T$ at different temperatures $T$.}
\label{drag_coeff}
\end{figure}

\begin{figure}[ht]
\includegraphics[width=2.0in,height=1.7in,angle=-90]{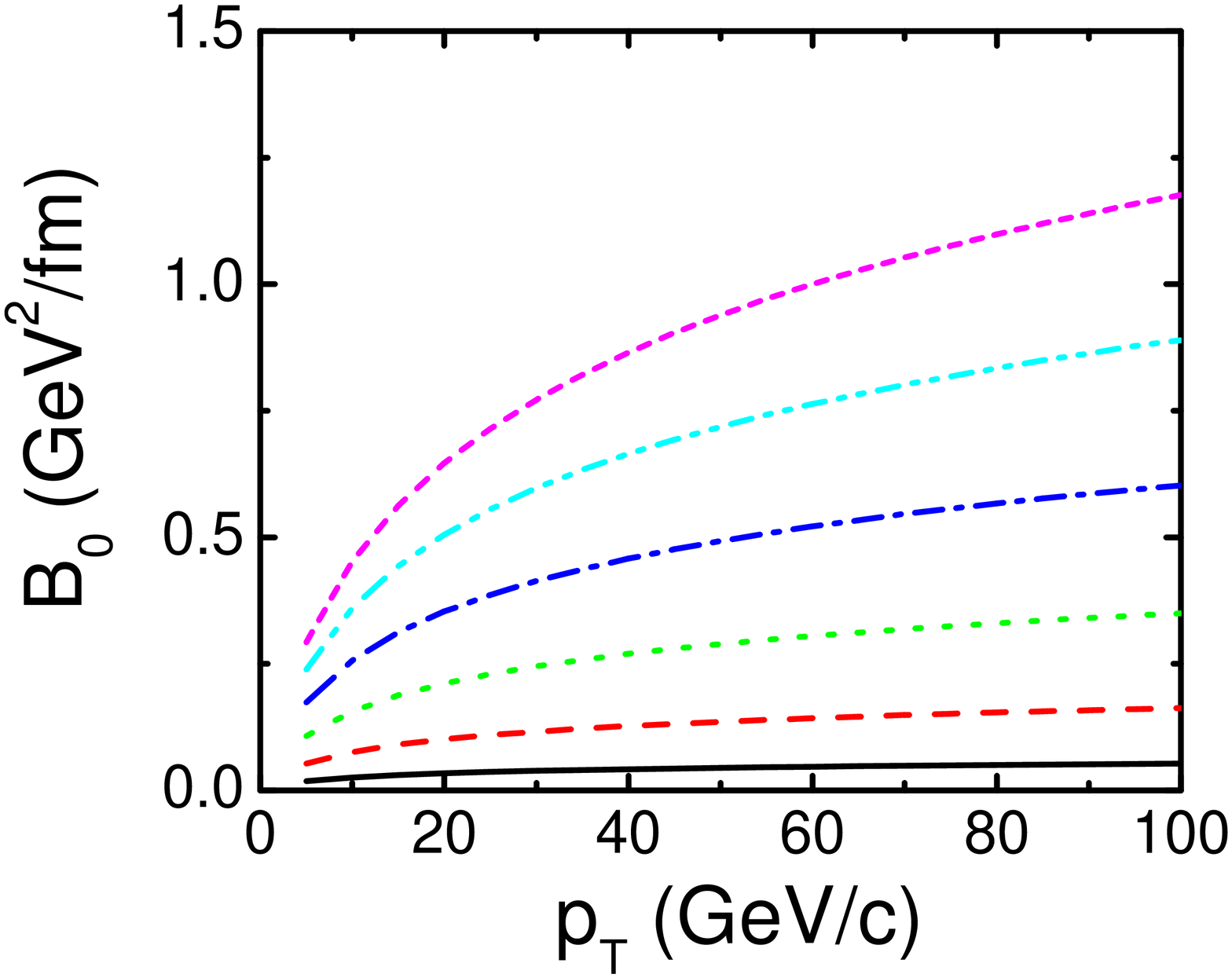}
\hspace{-0.5cm}
\includegraphics[width=2.0in,height=1.7in,angle=-90]{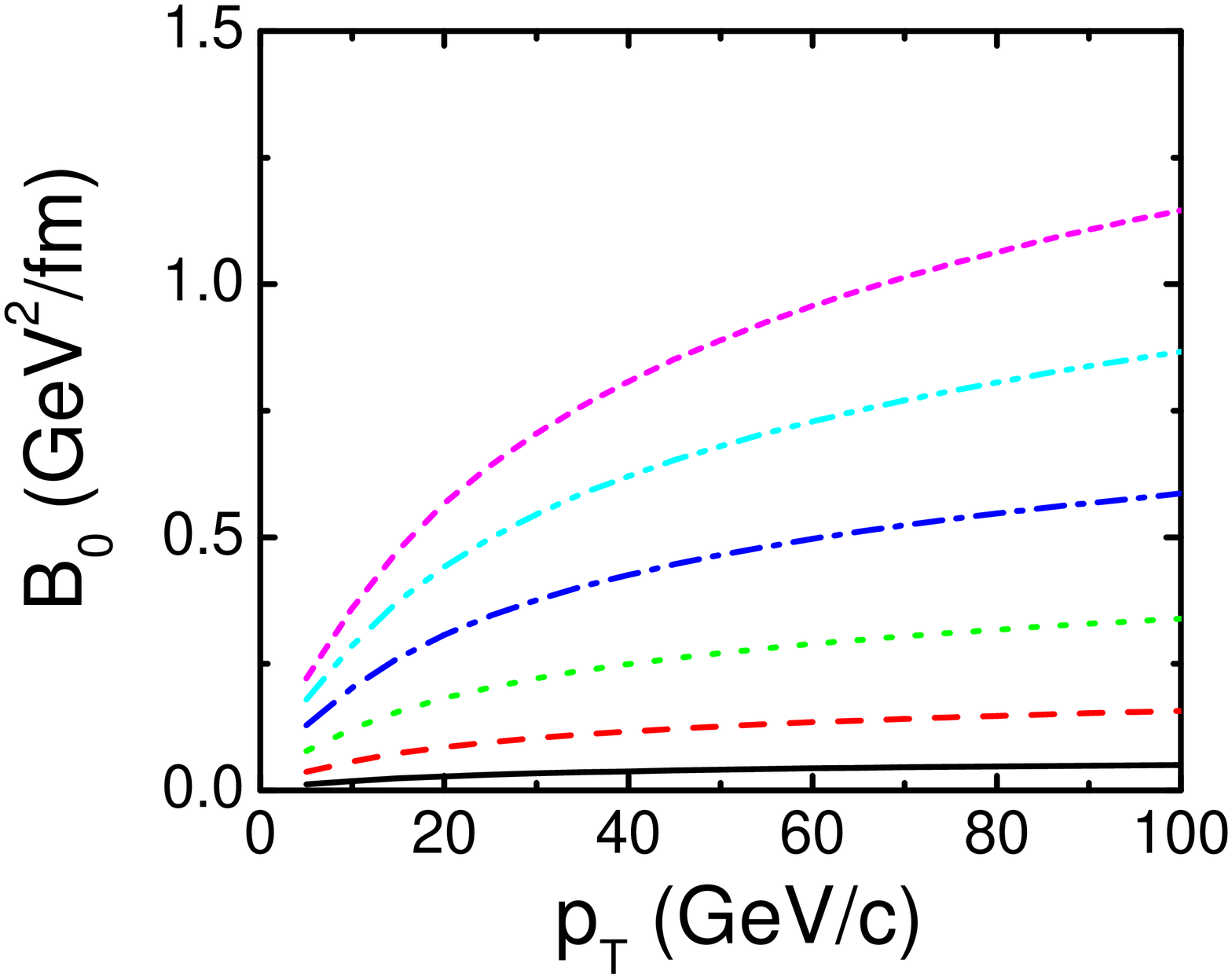}
\caption{(Color online) Transport coefficients $B_0$ of charm (left) and bottom (right) quarks in a QGP as functions of transverse momentum $p_T$ at different temperatures $T$.}
\label{diffu0_coeff}
\end{figure}

\begin{figure}[ht]
\includegraphics[width=2.0in,height=1.7in,angle=-90]{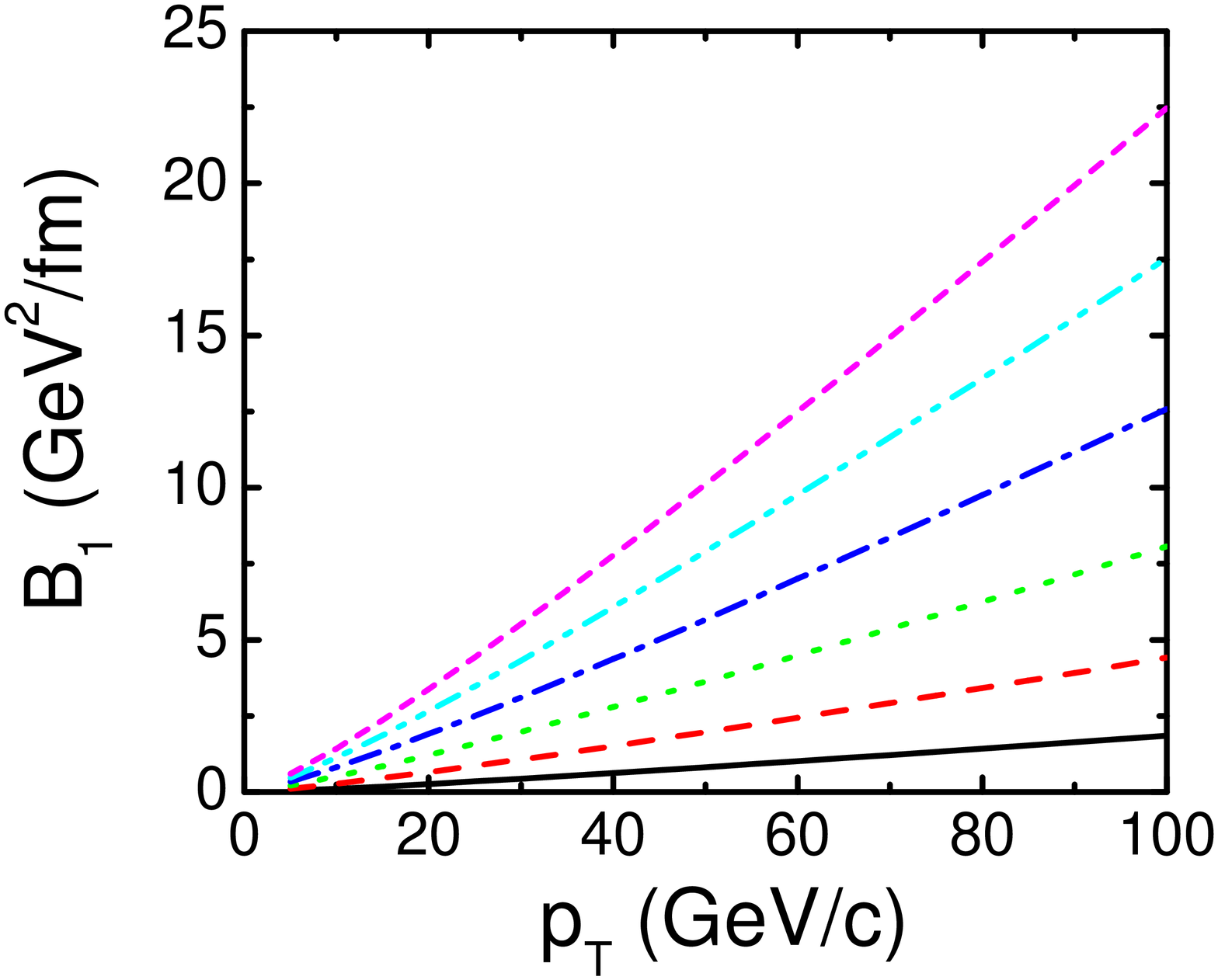}
\hspace{-0.5cm}
\includegraphics[width=2.0in,height=1.7in,angle=-90]{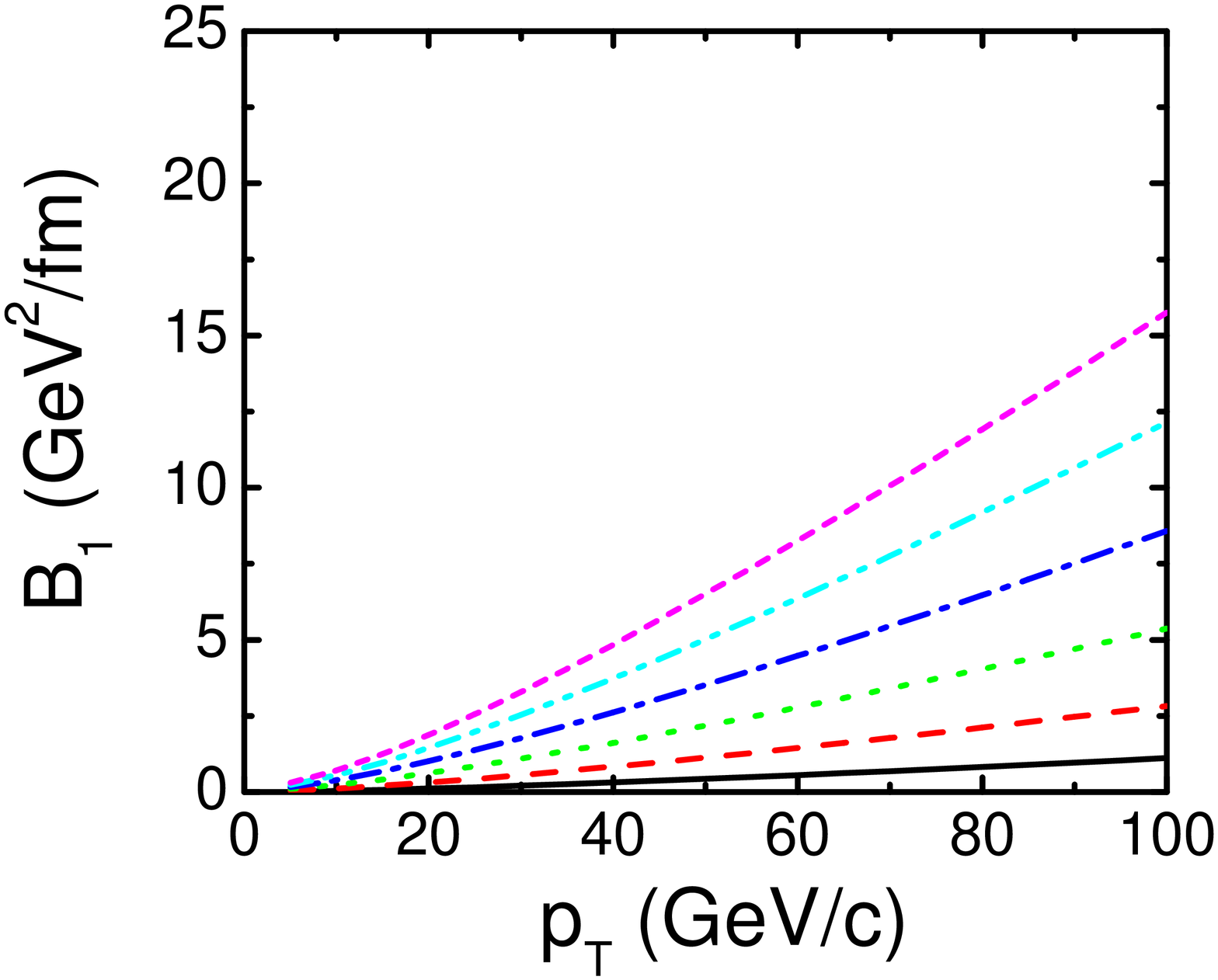}
\caption{(Color online) Transport coefficients $B_1$ of charm (left) and bottom (right) quarks in a QGP as functions of transverse momentum $p_T$ at different temperatures $T$.}
\label{diffu1_coeff}
\end{figure}

Using a fixed coupling constant $\alpha_s$ = 0.3 and masses $m_g=gT/\sqrt{2}$
for thermal gluons and $m_q=gT/\sqrt{6}$ for thermal quarks \cite{blazoit}, we
calculate the transport coefficients for both charm ($m_c$=1.2 GeV) and bottom
($m_b$=4.5 GeV) quarks numerically at leading order accuracy for the Compton
processes $Q+g(q)\to Q+g(q)$. We show the results in Figs.\ \ref{drag_coeff}, \ref{diffu0_coeff}, and \ref{diffu1_coeff}. The drag coefficients $\gamma$ decrease with increasing momentum of the incoming jets while diffusion coefficients behave the opposite way.

To study the dynamics of heavy quark in an expanding QGP medium, we focus on the degradation of transverse momentum of heavy quark jets, which can be taken into account using a Langevin equation \cite{hees1,moore} obtained from the Fokker-Planck equation in (\ref{fokkerplanck}) by calculating the first and second modes,
\begin{eqnarray}\label{langevin}
\Delta {x} &=& \frac{\vec p}{E}\Delta \tau,\nonumber\\
\Delta \vec{p} &=&-\gamma(T,\vec p+\Delta \vec p)\vec p \Delta \tau
+\Delta \vec W(T,\vec p+\Delta\vec p) \, .
\end{eqnarray}
Here $\Delta \vec W$ is the random force according to the normal distribution
\begin{eqnarray}
f_{rf}(\Delta \vec W) \sim \frac{1}{\sqrt{4\pi\Delta\tau}}
{\rm exp}\left[-\frac{\hat B_{ij}\Delta W^i\Delta W^j}{4\Delta\tau}\right]
\end{eqnarray}
and $\hat B_{ij}$ is the inverse of the diffusion-coefficient matrix $B_{ij}$ defined in Eq.\ (\ref{coefficient}).

Now we discuss the implementation of the conversion of light quarks (u,d,s)
and gluons into heavy quarks through the processes $g+g\to Q+\bar Q$ and
$g(q)+ Q\to Q+ g(q)$. The dynamics is governed by the rate equation
\begin{equation}
  \frac{dN^a}{d\tau} = -\sum_{b}\Gamma^{a\to b}(p_T)N^a
  +\sum_{c}\Gamma^{c\to a}(p_T)N^c \, .
\end{equation}
The collisional conversion widths can be estimated as
$\Gamma(p)=\langle 1\rangle$ in the notation defined in Eq.\ (\ref{average1}).

\begin{figure}[ht]
\includegraphics[width=2.0in,height=1.7in,angle=-90]{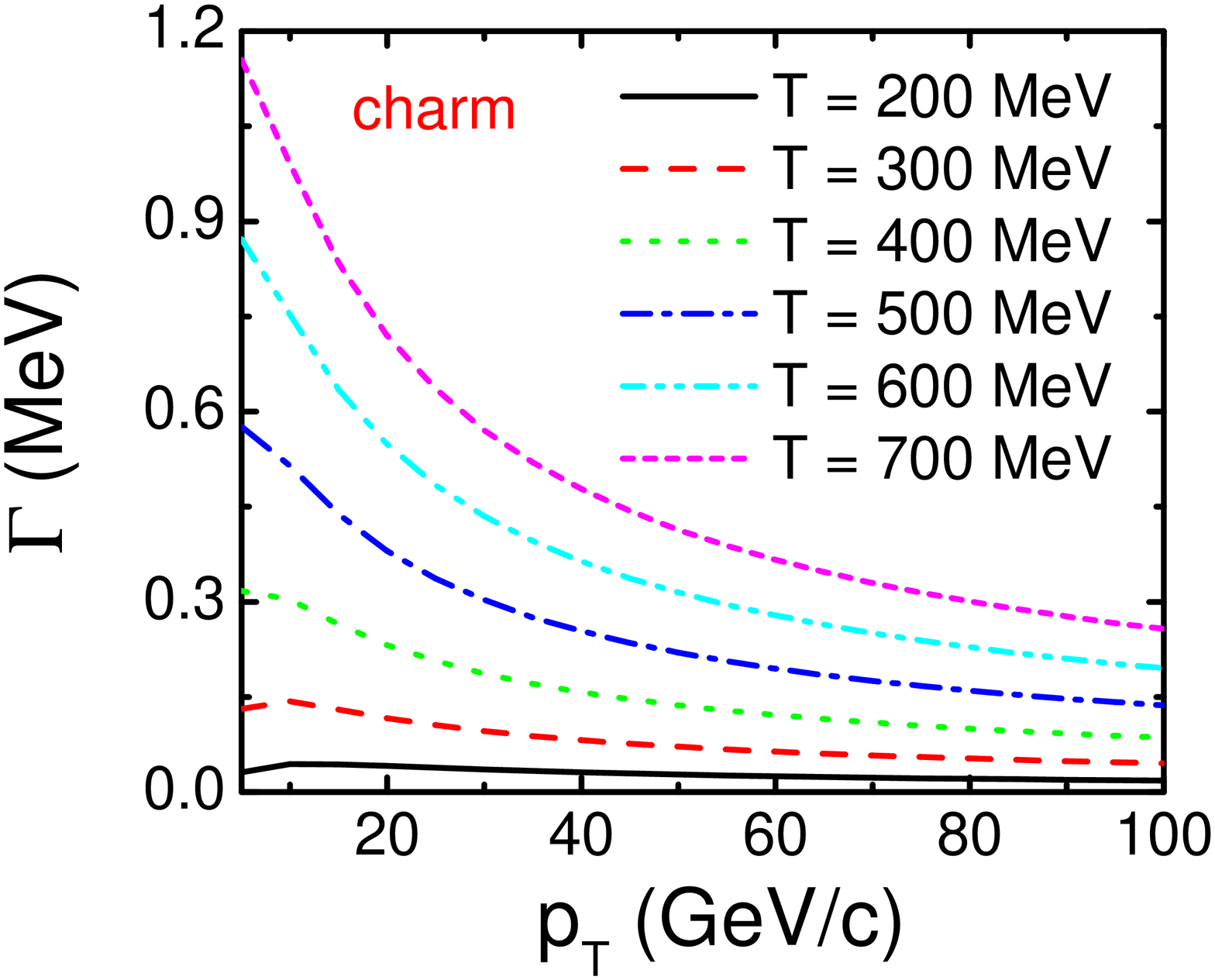}
\hspace{-0.5cm}
\includegraphics[width=2.0in,height=1.7in,angle=-90]{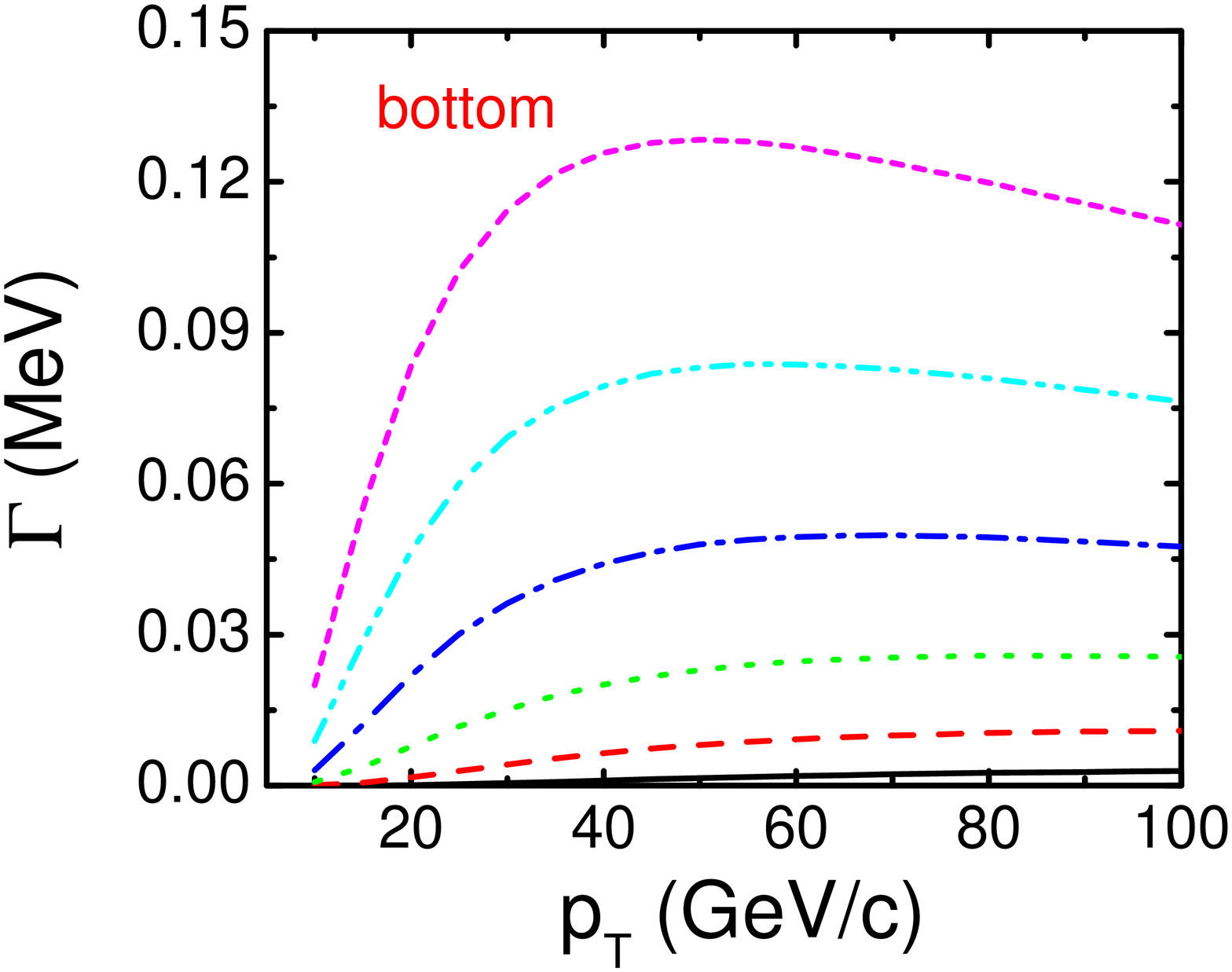}
\caption{(Color online) Conversion widths for charm (left panel) and bottom (right panel) in a QGP due to the process $g+g\to Q+\bar Q$ as functions of transverse momentum $p_T$ of the incoming gluon jets at different temperatures $T$.}
\label{heavy_width}
\end{figure}

\begin{figure}[ht]
\includegraphics[width=2.0in,height=1.7in,angle=-90]{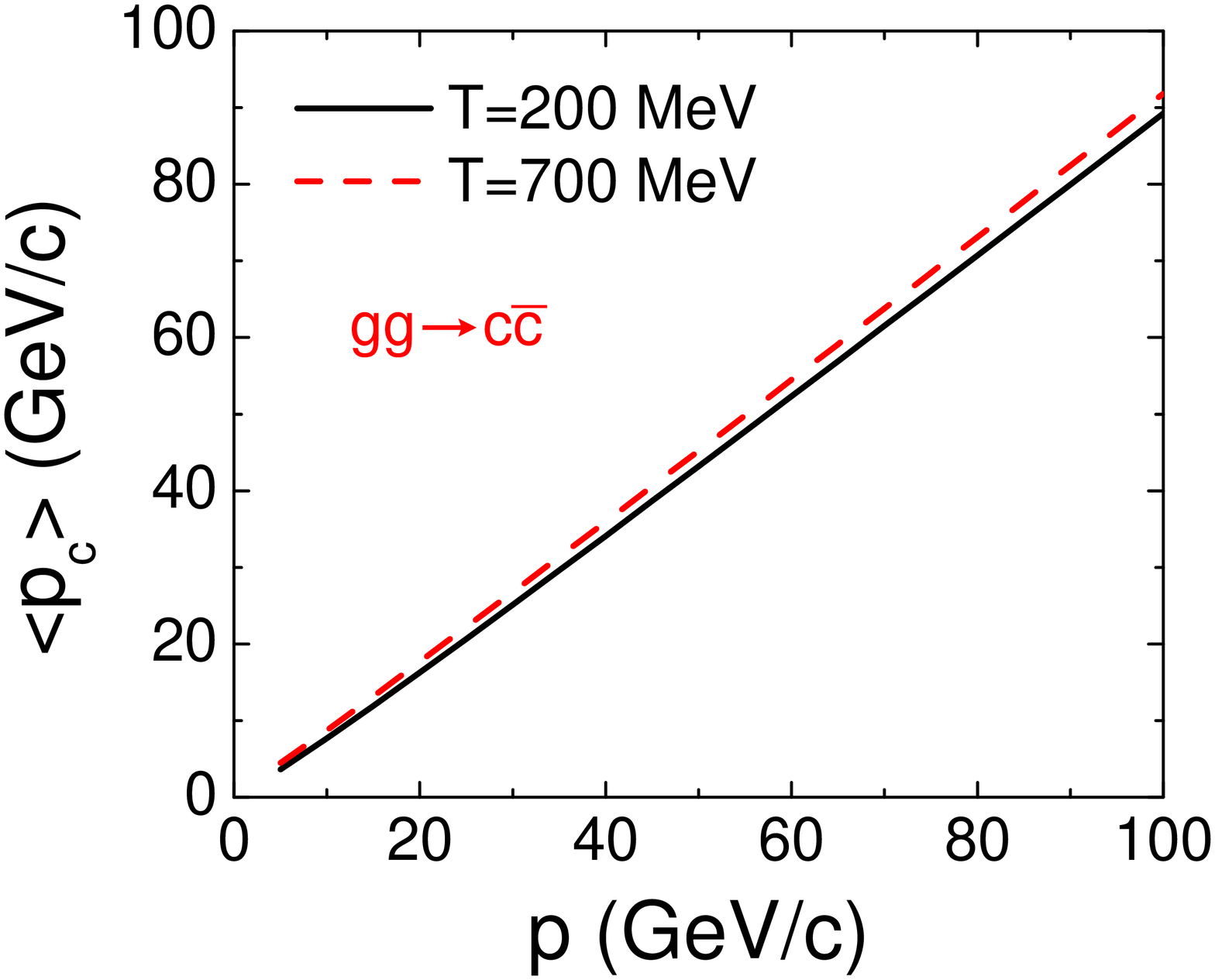}
\hspace{-0.2cm}
\includegraphics[width=2.0in,height=1.7in,angle=-90]{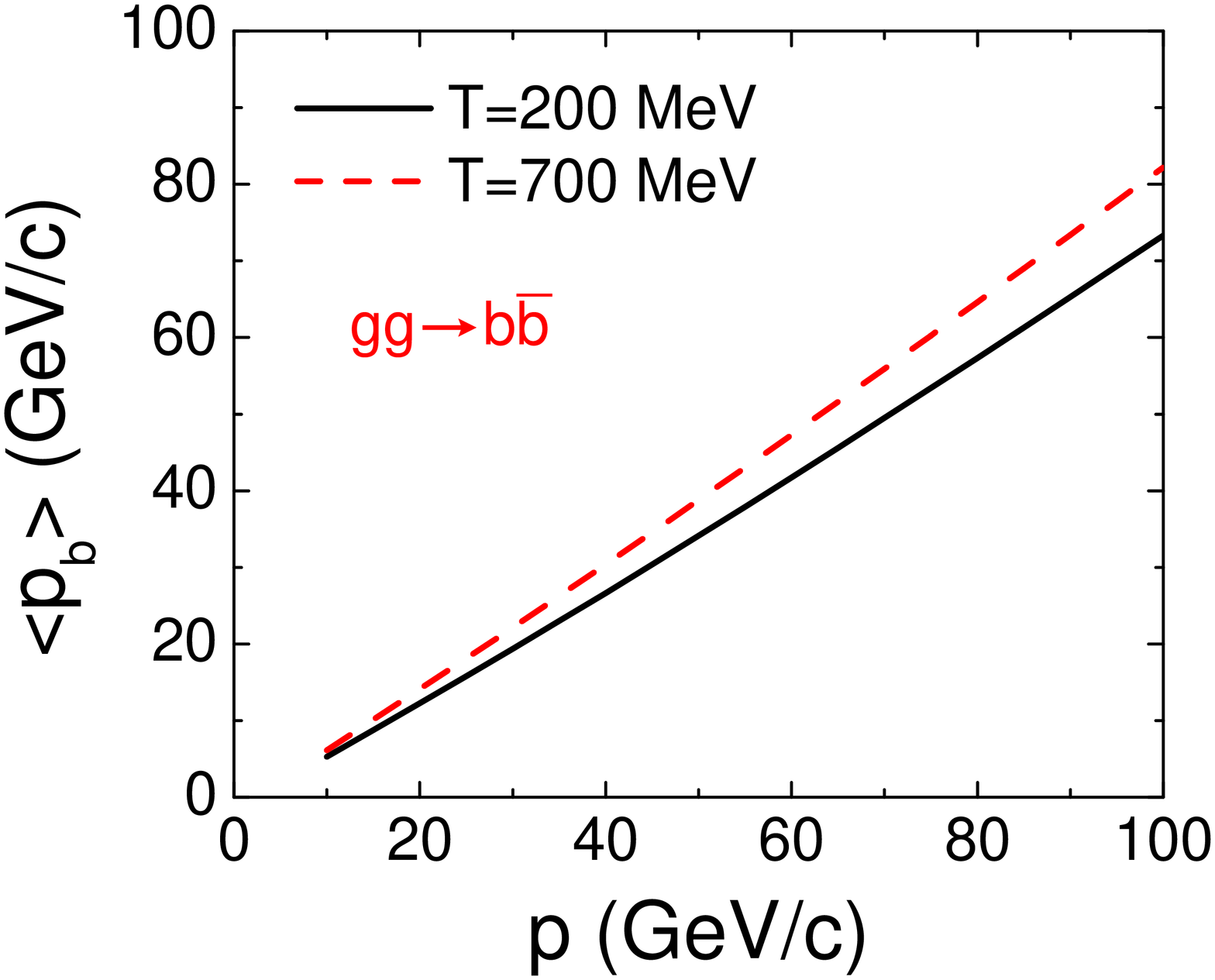}
\caption{(Color online) Average final state momentum of the heavy quark for charm (left panel) and bottom (right panel) after conversion from gluons with momentum $p$ for two different temperatures.}
\label{heavy_mom}
\end{figure}

In Fig.\ \ref{heavy_width}, we plot the width of a gluon jet converting to charm and bottom via the process $g+g\to Q+\bar Q$. We have introduced a constraint on the momentum of the outgoing heavy quark jet. We only count the heavy quark as a jet particle if it carries the larger momentum of the two final state partons.
The conversion will also cause effective energy loss since the final momentum
is usually smaller than the initial gluon jet momentum due to the mass threshold. This is different from the case of light flavor conversions which can occur with 100\% efficiency.
The resulting average momentum of the heavy quark after conversion from a
gluon, $p_Q = \langle p_Q\rangle/\langle 1\rangle$, as a function of the
incoming gluon momentum $p$ are shown in Fig.\ \ref{heavy_mom}.
$\langle p_Q \rangle$ has an almost linear dependence on $p$ only
depends very weakly on temperature.

As mentioned above there are two more elastic processes, $g+Q\to Q+g$ and
$q+Q\to q+Q$, which contribute to the enhancement of high-$p_T$ heavy quarks
by accelerating slowly moving heavy quarks from the medium. We use the
distribution function of heavy quarks obtained from perturbative QCD
calculations in Pb+Pb collisions at $\sqrt{s_{NN}}$=5.5 TeV (shown in the
next section) to estimate the conversion
width for a quark or gluon into a heavy quark via these Compton-like processes.
It turns out that their contribution to high-$p_T$ heavy quark production is
two orders of magnitude smaller than that from the gluon annihilation process
$g+g\to Q+\bar Q$ due to the mass mismatch in the initial state.
This is true when we require the momentum of the outgoing heavy quark to
be larger than half of the momentum of the incoming light flavor jet.
When this constraint is lifted a much larger conversion rate is
obtained, however, with a much smaller average momentum gain for the final
state heavy quark which seems irrelevant for the high-$p_T$ sector. Thus we
neglect the contributions from Compton-like elastic processes in the following.

In this work we follow the procedure outlined for light quark and gluons in
\cite{weiliu1}. The leading order matrix elements are scaled by a factor
$K$ = 4 which was needed to reproduce the observed $p/\pi^+$ and
$\bar p/\pi^-$ ratios. Ultimately, however, the drag and diffusion coefficients
have to be extracted from data and compared to a variety of perturbative
and non-perturbative predictions. A large $K$-factor as used here, if found compatible with data, might rather hint toward a much stronger jet-medium coupling
than expected from perturbation theory.

\section{Heavy Quark Conversions at RHIC and LHC}
\label{propagate}

Now we turn to charm and bottom production in Au+Au collisions at RHIC with a center-of-mass energy $\sqrt{s_{NN}}=200$ GeV. The initial $p_T$ spectra of charm and bottom quarks at mid-rapidity are taken from Ref.\ \cite{weiliu}. Both are obtained by multiplying the heavy quark $p_T$-spectra from $p+p$ collisions at the same energy by the number of binary collisions ($\sim 960$) in central Au+Au collisions.

\begin{figure}[ht]
\includegraphics[width=2.0in,height=1.7in,angle=-90]{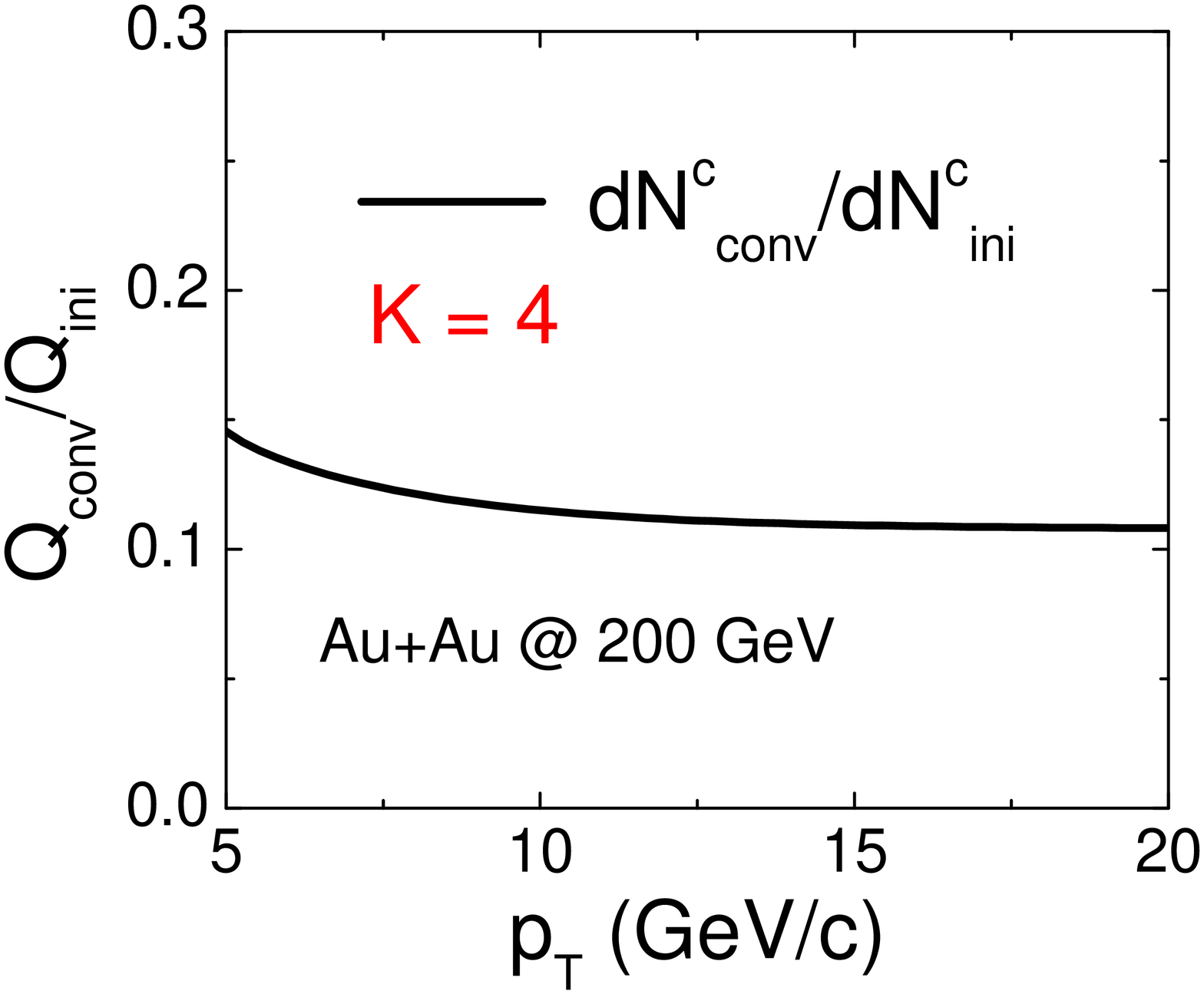}
\hspace{-0.2cm}
\includegraphics[width=2.0in,height=1.7in,angle=-90]{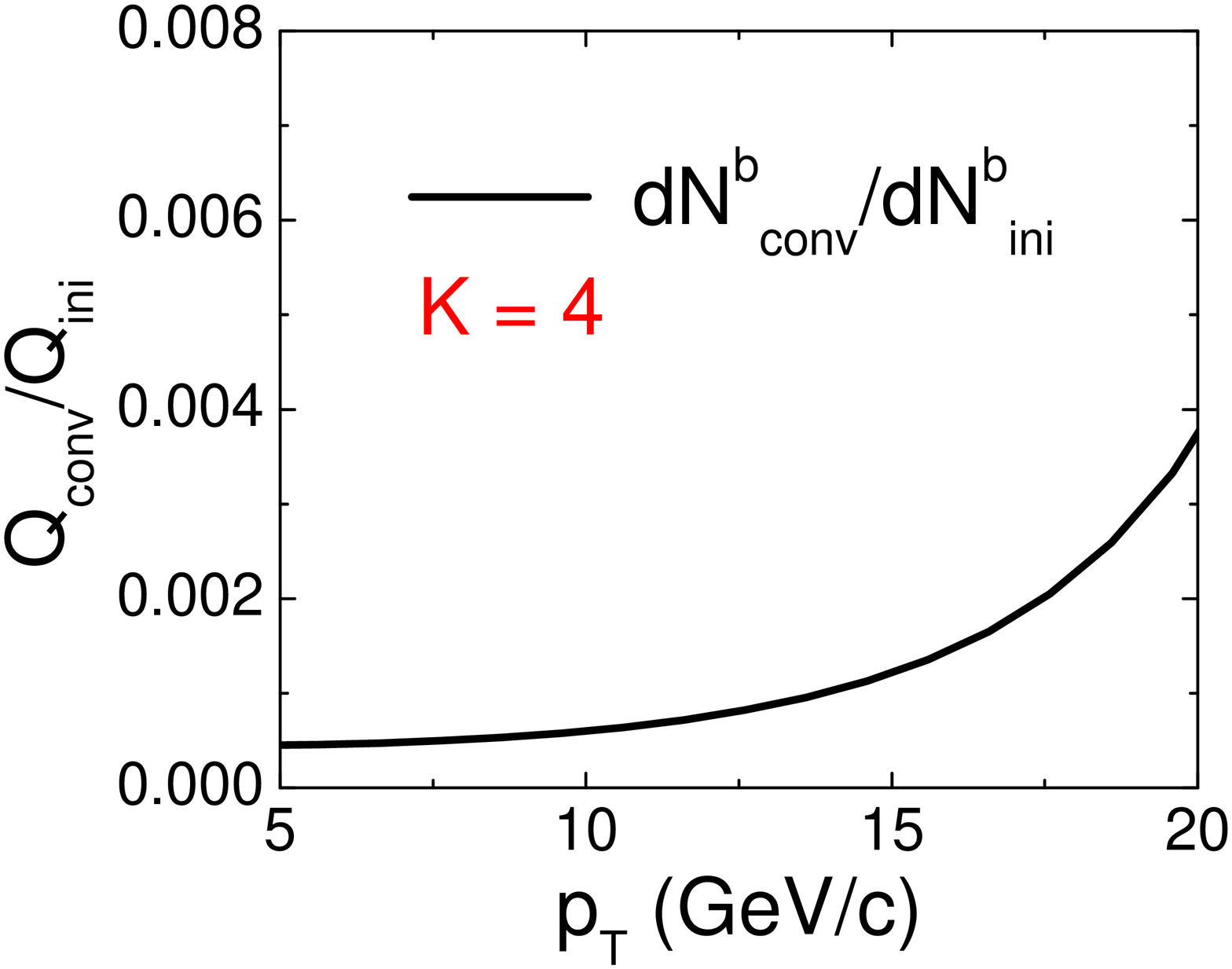}
\caption{(Color online) The ratio of heavy quarks from jet conversions
to heavy quarks from initial production with energy loss included for charm (left panel) and bottom (right panel) as functions of transverse momentum $p_T$ in Au+Au collisions at
$\sqrt{s_{NN}}$ = 200 GeV.}
\label{heavy_conv}
\end{figure}

For the dynamics of the fireball, we assume that it evolves boost invariantly
in the longitudinal direction but with an accelerated transverse expansion.
Specifically, its volume expands in the proper time $\tau$ according to
$V(\tau)=\pi R(\tau)^2\tau$, where $R(\tau)=R_0+a/2(\tau-\tau_0)^2$
is the transverse radius with an initial value $R_0$=7 fm,
$\tau_0$=0.6 fm is the QGP formation time, and $a=0.1c^2$/fm is the
transverse acceleration \cite{weiliu, weiliu1, lwchen}. With an initial temperature $T_i=350$ MeV and using thermal masses for quarks and gluons, this model gives a total transverse energy comparable to that measured in experiments. The time dependence of the temperature is then obtained from entropy conservation, and the critical temperature $T_c=175$ MeV is reached
at proper time $\tau_c\sim 5$ fm.

To solve Eq.\ \ref{langevin}, we follow the test particle methods introduced in Ref.\ \cite{weiliu1} for the propagation of light flavored jets and heavy quarks in the QGP medium. The light flavors can convert into each other or into heavy quarks according to the probabilities discussed above in \cite{weiliu2}.
The conversion of heavy quarks into light flavors is suppressed and not
taken into account, except for the process $Qg(q)\to g(q)Q$ which is
included in the drag coefficient.
In Fig.\ \ref{heavy_conv} we show the ratio of heavy quarks obtained from jet conversions to that obtained from initial hard collisions with energy loss included. The jet conversions contribute about 10\% to the spectrum of charm quarks at
high $p_T$, and only 0.3\% to that of bottom quarks in central Au+Au collisions at RHIC.

\begin{figure}[ht]
\includegraphics[width=2.0in,height=1.7in,angle=-90]{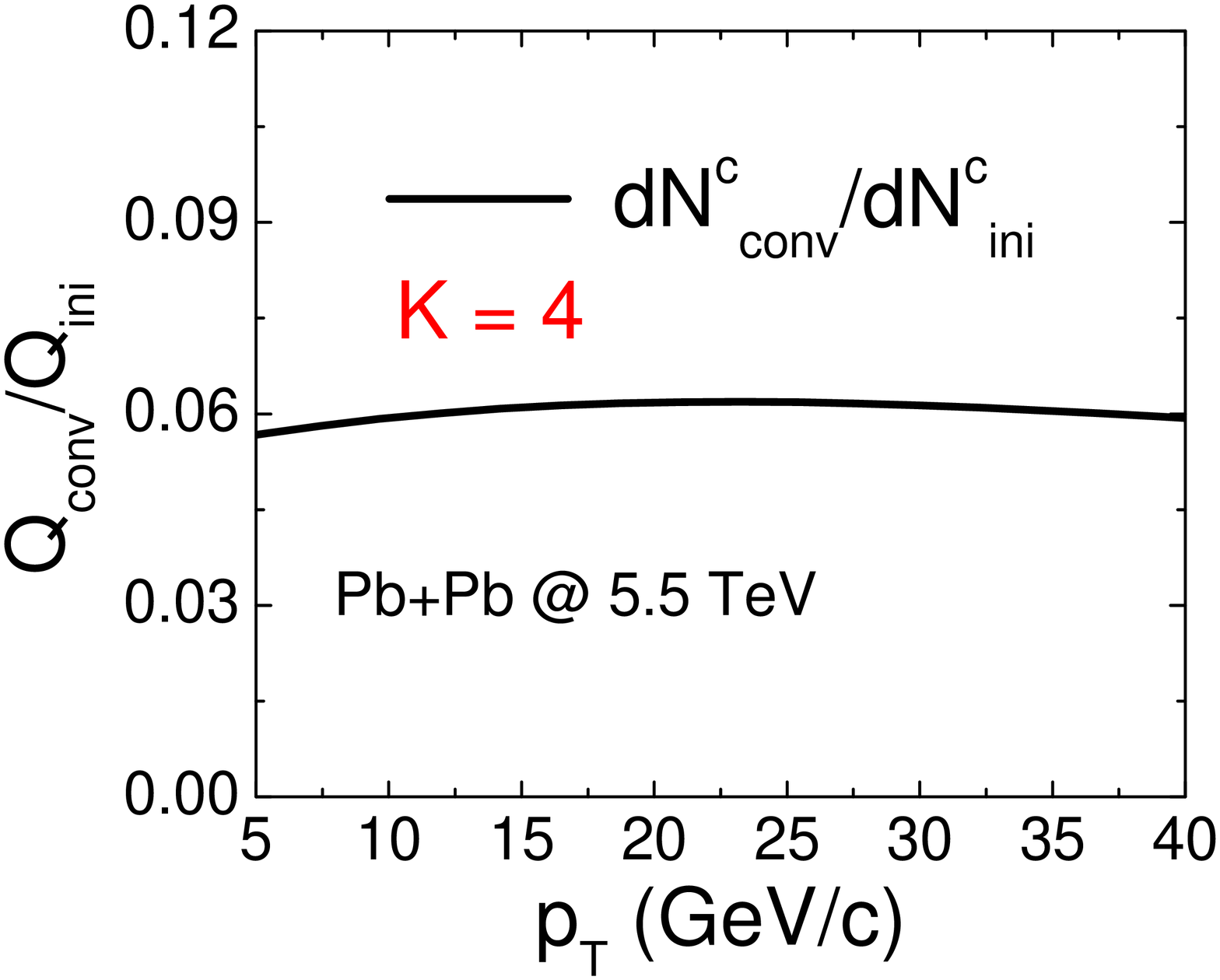}
\hspace{-0.2cm}
\includegraphics[width=2.0in,height=1.7in,angle=-90]{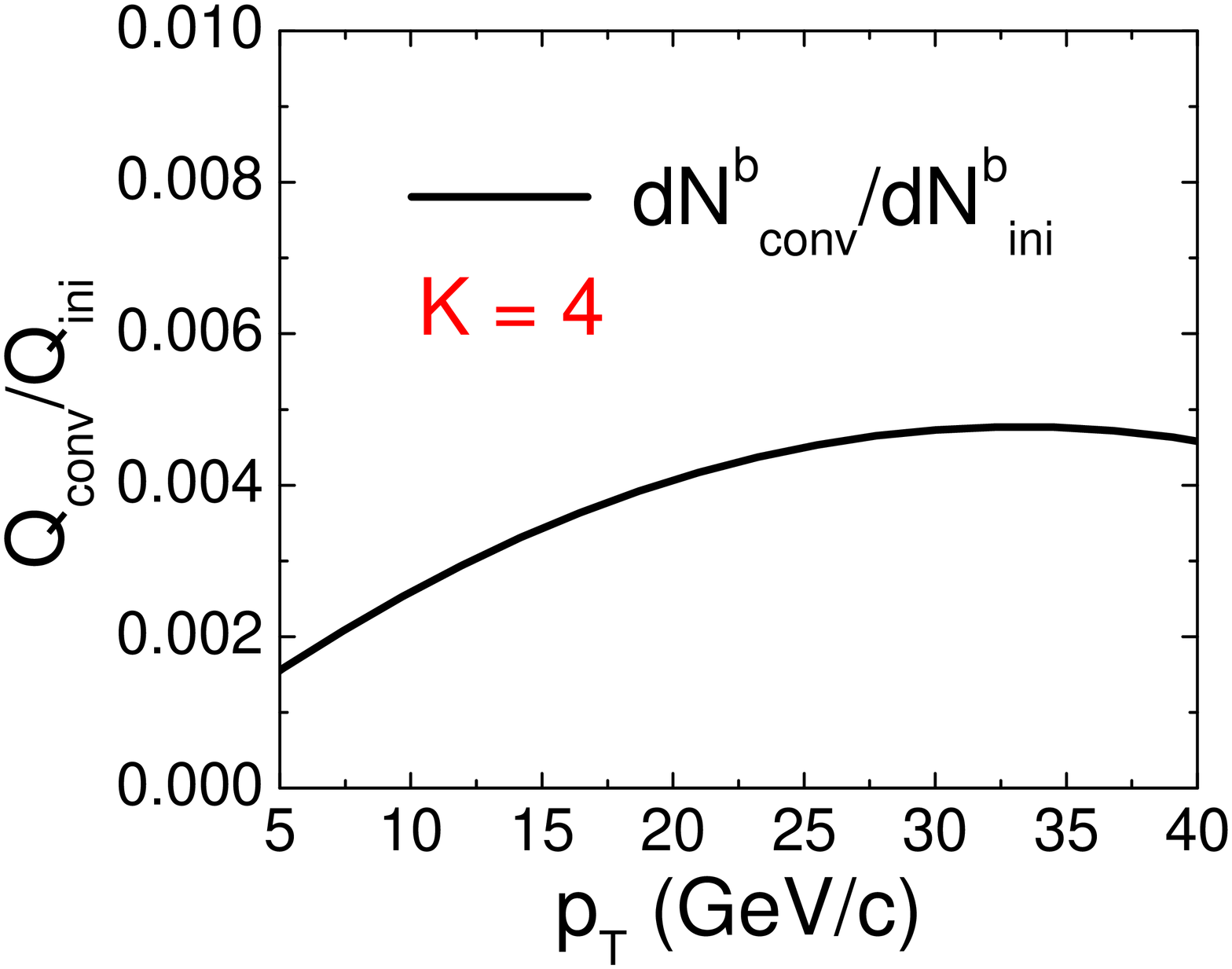}
\caption{(Color online) The ratio of heavy quark spectrum from initial production to that from jet conversions for charm (left panel) and bottom (right panel) as functions of transverse momentum in Pb+Pb collisions at$\sqrt{s_{NN}}$ = 5.5 TeV.}
\label{heavy_con_lhc}
\end{figure}

For LHC energies we take the initial $p_T$ spectra of charm and bottom quarks
at midrapidity from the perturbative calculation in Refs.\ \cite{vogt,carrer}
multiplied by the number of binary collisions
($\langle N_{coll}\rangle\approx$ 1700 \cite{david}). The spectra
are parametrized as
\begin{eqnarray}
\frac{dN_c}{d^2 p_T}&=&2497\left(1+\frac{p_T}{1.95}\right)^{-5.5}\left( p_T\left[1+\left(\frac{4}{0.1+p_T}\right)^2\right]\right)^{-1}\nonumber\\
\frac{dN_b}{d^2 p_T}&=&38\left(1+\frac{p_T}{2.0}\right)^{-5.6} \left[1+\left(\frac{10}{0.8+p_T}\right)^3\right]^{-1},
\end{eqnarray}
The parameters of the fireball are taken from Ref.\ \cite{zhang} where thermal charm production at LHC was studied. Specifically, we choose initial proper time $\tau_0=0.2$ fm/$c$ for the formation of the equilibrated QGP with initial temperature $T_0=700$ MeV.
In Fig.\ \ref{heavy_con_lhc}, we show the ratio of heavy quark yields from jet conversions to that from initial hard collisions with energy loss included.
The ratios are similar in magnitude magnitude to what has been obtained for
RHIC.

We can now answer the two questions that we have posed at the beginning.
First, the promotion of heavy quarks from the medium to high $p_T$ is not
an effective mechanism for the production of high-$p_T$ charm and bottom
quarks. This is compatible with the experimental observations \cite{adare}.
Note that we have used the same framework and $K$-factors which led to the
prediction of a sizable strange quark excess at high-$p_T$ at RHIC. On the
other hand, we have found that the corresponding process for charm quarks at
LHC is not efficient. This can be traced back to the relatively large amount
of initial charm in jets produced at LHC and the missing chemical
equilibration of charm in the QGP. Note that some authors have
argued recently that thermal charm production might occur at the
temperatures reached at LHC \cite{zhang}. We have chosen not to include this
scenario here.

\section{Discussion}\label{summary}

We have studied the impact of jet conversions on the production of high-$p_T$ heavy quarks in heavy ion collisions. Based on leading order perturbative QCD, we calculated transport coefficients of heavy quarks, and studied their propagation in an expanding QGP medium.
We find that the dominant conversion process is pair production off gluon
jets in the medium, $g+g\to Q+\bar Q$. However, the contribution of jet conversions to the total yields of heavy quarks at high $p_T$ is rather small both
at RHIC and LHC.
This is consistent with measurements of single electrons at moderate values
of $p_T$ at RHIC. On the other hand, it implies that high-$p_T$
charm quarks at LHC as a probe for the mean free path in the medium
might not be as useful as strangeness is at RHIC energies.

\begin{acknowledgments}
We wish to thank Ralf Rapp and Che-Ming Ko for helpful discussions. This work was supported by RIKEN/BNL, DOE grant DE-AC02-98CH10886, and the Texas A\&M College of Science.
\end{acknowledgments}


\begin{thebibliography}{99}

\bibitem{Baier:1996kr}
  R.~Baier, Y.~L.~Dokshitzer, A.~H.~Mueller, S.~Peigne and D.~Schiff,
  Nucl.\ Phys.\  B {\bf 483}, 291 (1997);
  B.~G.~Zakharov,
  JETP Lett.\  {\bf 65}, 615 (1997).

\bibitem{wang} X.\ N.\ Wang, Phys.\ Lett.\ B {\bf 579}, 299 (2004).
\bibitem{gyulassy}M.\ Gyulassy, P.\ L\'evai, and I.\ Vitev,
Phys.\ Rev.\ Lett.\ {\bf 85}, 5535 (2001).
\bibitem{wiedemann} U.\ A.\ Wiedemann, Nucl.\ Phys.\ B {\bf 588}, 303 (2000).
\bibitem{adcox} A.\ Adcox {\it et al.} (PHENIX Collaboration), Phys.\ Rev.\
Lett.\ {\bf 88}, 022301 (2002).
\bibitem{adler1} C.\ Adler {\it et al.} (STAR Collaboration), Phys.\ Rev.\
Lett.\ {\bf 89}, 202301 (2002); {\bf 90}, 082302 (2002).
\bibitem{adare} A.\ Adare {\it et al.} (PHENIX Collaboration), Phys.\ Rev.\ Lett.\ 96, 032301 (2006); Phys.\ Rev.\ Lett.\ 98, 172301 (2007); B.\ I.\ Abelev {\it et al.} (STAR Collaboration), Phys.\ Rev.\ Lett.\ 98, 192301, 2007.
\bibitem{mustafa} M.\ G.\ Mustafa, Phys.\ Rev.\ C {\bf 72}, 014905 (2005);
M.\ G.\ Mustafa {\it et al.}
\bibitem{wicks} S.\ Wicks {\it et al.}, {\it Preprint}
nucl-th/0512076.
\bibitem{djordjevic1} M.\ Djordjevic, M.\ Gyulassy, and S.\ Wicks,
Phys.\ Rev.\ Lett.\ {\bf 94}, 112301 (2005).
\bibitem{hees} H.\ van Hees and R.\ Rapp, Phys.\ Rev.\ C {\bf 71}, 034907 (2005); H.\ van Hees, V.\ Greco, and R.\ Rapp, nucl-th/0508055.
\bibitem{weiliu1}W.\ Liu, C.\ M.\ Ko, and B.\ W.\ Zhang, Phys.\ Rev.\ C75, 051901  (2007); J.\ Mod.\ Phys.\ E.\ 16, 1930 (2007).
\bibitem{Fries:2002kt}
  R.~J.~Fries, B.~Muller and D.~K.~Srivastava,
  Phys.\ Rev.\ Lett.\  {\bf 90}, 132301 (2003);
  D.~K.~Srivastava, C.~Gale and R.~J.~Fries,
  Phys.\ Rev.\  C {\bf 67}, 034903 (2003);
  S.~Turbide, C.~Gale, D.~K.~Srivastava and R.~J.~Fries,
  Phys.\ Rev.\  C {\bf 74}, 014903 (2006).
\bibitem{adams} J. Adams {\it et al.} (STAR Collaboration), {\it Preprint} nucl-ex/0606003.
\bibitem{weiliu2}W.\ Liu and R.\ J.\ Fries, {\it Preprint} 0801.0453 [nucl-th].
\bibitem{Sapeta:2007ad}
  S.~Sapeta and U.~A.~Wiedemann,
  {\it Preprint} arXiv:0707.3494 [hep-ph].
\bibitem{benjamin} B.\ Svetitsky, Phys.\ Rev.\ D {\bf 37}, 2484 (1988).
\bibitem{blazoit} J.\ P.\ Blaizot and E.\ Iancu, Phys.\ Rep.\ {\bf 359}, 355 (2002).
\bibitem{hees1} H.\ van Hees, V.\ Greco and R.\ Rapp, nucl-th/0508055.
\bibitem{moore} G.\ D.\ Moore and D.\ Teaney, Phys.\ Rev.\ C {\bf 71}, 064904 (2005).
\bibitem{weiliu} W.\ Liu and C.\ M.\ Ko, {\it Preprint} nucl-th/0603004; J.\ Phys.\ G.\ 34, s775 (2007); C.\ M.\ Ko and W.\ Liu, Nucl.\ Phys.\ A {\bf 783}, 233  (2007).
\bibitem{lwchen} L.W. Chen, V. Greco, C.M. Ko, S.H. Lee, and W. Liu, Phys. Lett. B {\bf 601}, 34 (2004).
\bibitem{vogt} J.\ Baines {\it et al.}, {\it Preprint} arXiv: hep-ph/0003142.
\bibitem{carrer} N. Carrer and A. Dainese, {\it Preprint} arXiv: hep-ph/0311255.
\bibitem{david} D.\ d'Enterria, {\it Preprint} nucl-ex/0302016.
\bibitem{zhang} B.~W.~Zhang, C.~M.~Ko and W.~Liu,
  Phys.\ Rev.\  C {\bf 77}, 024901 (2008).


\end{thebibliography}
\end{document}